\documentclass[prb,twocolumn,superscriptaddress,showkeys,floatfix]{revtex4-2}

\usepackage{graphicx}
\usepackage{ulem}
\usepackage{bm}
\usepackage{xcolor}
\usepackage{url}
\usepackage{bookmark}
\usepackage{amsmath}

\newcommand{\be}{\begin{equation}}
\newcommand{\ee}{\end{equation}}

\def\bea{\begin{eqnarray}}
\def\eea{\end{eqnarray}}

\def\vq{{\bf q}}

\def\vk{{\bf k}}

\begin{document}

\title{Out-of-plane bond-order phase, superconductivity, and their competition in the $t$-$J_\parallel$-$J_\perp$ model:
Possible implications for bilayer nickelates}

\author{Mat\'{\i}as Bejas}
\affiliation{Facultad de Ciencias Exactas, Ingenier\'{\i}a y Agrimensura and Instituto de F\'{\i}sica Rosario (UNR-CONICET), Avenida Pellegrini 250, 2000 Rosario, Argentina}

\author{Xianxin Wu}
\affiliation{Institute of Theoretical Physics, Chinese Academy of Sciences, Beijing, China}

\author{Debmalya Chakraborty}
\affiliation{Max Planck Institute for the Physics of Complex Systems, N\"othnitzer Stra\ss e 38, 01187, Dresden, Germany}
\affiliation{Department of Physics, Birla Institute of Technology and Science - Pilani, K. K. Birla Goa Campus, NH-17B, Zuarinagar, Sancoale, Goa- 403726, India}

\author{Andreas P. Schnyder}
\affiliation{Max Planck Institute for Solid State Research, Heisenbergstrasse 1, 70569 Stuttgart, Germany}

\author{Andr\'es Greco}
\affiliation{Facultad de Ciencias Exactas, Ingenier\'{\i}a y Agrimensura and Instituto de F\'{\i}sica Rosario (UNR-CONICET), Avenida Pellegrini 250, 2000 Rosario, Argentina}
\affiliation{Max Planck Institute for Solid State Research, Heisenbergstrasse 1, 70569 Stuttgart, Germany}

\date{\today}

\begin{abstract}
Almost four decades of intense research have been invested 
to study the physics of high-T$_c$ cuprate superconductors.
The recent discovery of high-T$_c$ superconductivity in pressurized bilayer nickelates and its potential similarities with cuprate superconductors may open a new window to understand this long-standing problem.
We have studied the proposed bilayer $t$-$J_\parallel$-$J_\perp$ model [where $J_\parallel$ ($J_\perp$) is the in-plane (out-of-plane) magnetic exchange] in a large-$N$ approach on the basis of the path integral representation for Hubbard operators, which allows to obtain results at mean-field and beyond mean-field level.
We find that $J_\perp$ is a candidate for triggering high superconducting $T_c$ values at quarter filling (hole doping $\delta=0.5$) of the $d_{x^2-y^2}$ orbitals.
Beyond mean-field level, we find a new phase, an out-of-plane bond-order phase (z-BOP), triggered also by $J_\perp$. z-BOP  develops below a critical temperature which decreases with increasing doping and vanishes at a quantum critical point below quarter filling.
The occurrence of this phase and its competition with superconductivity leads to a superconducting dome-shaped behavior as a function of doping and as a function of $J_\perp$.
Qualitative comparisons with the physics of cuprates and the recent literature on the new pressurized nickelates are given along the paper.
\end{abstract}


\maketitle

\section{introduction}

The recent discovery of superconductivity in the bilayer nickelates La$_3$Ni$_2$O$_7$ at high critical temperature $T_c \sim 80$ K \cite{sun23,hou23} under moderate pressure of about $14$ GPa has motivated a huge experimental \cite{yang23e2,kakoi24,sakakibara24,li24,chen24,puphal23,zhou24,liu23,zhang23,li24p,chen24p,zhu24p,xie24,chen24pp,dan24new,abadi24,wang24pp,zhang24pp,meng2024} and theoretical
\cite{yang23,luo23,ZhangY2023,Lechermann2023,Sakakibara2023,gu23,liao23,qin24,yang23t,qin23,zheng23,lu23,luo23p,lu24,qu24,oh24t,Jiang_2024,shen23,wu24,liu23t,yang23p,pan23,qu23p,tian24,wang24ppp,zhan24ppp,oh2024,tian2024,liao2024} interest.
We have cited only part of the vast existing literature; see also \cite{wang2024normal} for a review.
Trilayer pressurized La$_4$Ni$_3$O$_{10}$ nickelates have been also studied, and  a lower value of $T_c$ ($T_c \sim 30$ K) was obtained \cite{sakakibara24,kakoi24,li24p,zhu24p,zhang24pp}.  

The experiments are not straightforward, because they are done at high pressure, such that the nature of the observed superconductivity has become the subject of a debate \cite{hepting23}.
For instance, in bilayer nickelates the volume fraction for superconductivity was primarily estimated experimentally to be of the order of 1\% \cite{zhou24}, and while some experiments reported zero residual resistance \cite{zhang23,wang24pp,li24pppp,hou23} others did not \cite{zhou24,sun23,puphal23}. 
Based on that situation, it was assumed that a filamentary superconductivity occurs in this material \cite{puphal23,zhou24}.
Instead, trilayer nickelates shows about 80\% of volume fraction for superconductivity \cite{zhu24p}.
However, recently the situation seems to have substantially changed.
A volume fraction for superconductivity of about 50\% and zero residual resistance were reported \cite{li24pppp} in bilayer nickelates.  The substitutions of Pr for La are found to 
effectively inhibit the intergrowth of different phases and result in a nearly pure bilayer structure, leading to bulk high T$_c$ superconductivity in La$_2$PrNi$_2$O$_7$ \cite{wang2024bulk}. In addition, the superconducting Meissner effect was very recently detected \cite{wen2024probing}.
These findings, together with the high superconducting volume fraction in trilayer nickelates,  might locate the new pressurized nickelates inside the category of bulk high-$T_c$ superconductors as cuprates.
In addition, in Ref. \cite{li24pppp} 
the existence of a dome for $T_c$ as a function of pressure was reported.
That is, $T_c$ increases with pressure reaching the maximum at 18 GPa, and for larger pressure $T_c$ decreases.

For almost four decades cuprate superconductors were extensively investigated  motivated by their high value of $T_c$ and their anomalous properties, such as the pseudogap, strange metal phase, charge orders, and other features \cite{keimer15,timusk99,Hayden24,Proust19,Pepin20}.
It is widely considered that cuprate superconductors are prototype materials for understanding correlated systems \cite{Lee06a}.
Through the years many systems had been studied such as, for instance, organic materials \cite{mckenzie97}, cobaltates \cite{takada03}, iron superconductors \cite{scalapino12}, and more recently the kagome materials \cite{ortiz19}.
In all of them the possible contact with the physics of cuprates is discussed, although their $T_c$ is much smaller. 

Since there are close similarities between nickelates and cuprate superconductors \cite{Norman20}
it is probable that the discovery of superconductivity in pressurized nickelates opens a new window to investigate the topic from another perspective.
Several effective models have been proposed for nickelates, which run from multiorbitals ($d_{z^2}$ and $d_{x^2-y^2}$) \cite{qin23,yang23,gu23,luo23,lu23,liao23,shen23,wu24,yang23p,qu23p,zheng23} to only one active quarter filling orbital $d_{x^2-y^2}$ \cite{lu24,qu24,yang23t} models.

Motivated by the analogies with high-T$_c$ cuprates and strong coupling between NiO$_2$ layers through the inner apical oxygen, the strongly correlated $t$-$J_\parallel$-$J_\perp$ \cite{lu24,qu24,oh24t} model was proposed as a candidate
to describe the physics of bilayer nickelates La$_3$Ni$_2$O$_7$,
where $J_\parallel$ and $J_\perp$ are the in-plane and out-of-plane magnetic exchanges, respectively. In this model only the $d_{x^2-y^2}$ orbital plays an active role. Recent reports \cite{liu2024,liu2024p} suggest the dominant contribution of the $d_{x^2-y^2}$ orbitals.
The model consists of the usual $t$-$t'$-$J_\parallel$ model for each plane coupled by an out-of-plane $J_\perp$ between planes.
In addition, a small electron hopping $t_\perp$ between planes is expected \cite{lu24,luo23}.
This model was also discussed in the context of the slave boson approach at mean-field level \cite{lu24}, the density matrix renormalization group, and infinite projected entangled-pair states \cite{qu24}.

In the present paper we study the $t$-$J_\parallel$-$J_\perp$ model in a large-$N$ approximation on the basis of the path integral representation \cite{foussats99,foussats04} for Hubbard operators \cite{hubbard63}, and perform calculations at mean-field and beyond mean-field level.
We find tendencies to an out-of-plane $s$-wave superconductivity triggered by $J_\perp$; only if $J_\perp$ is larger than a certain minimum value.
Notably, we also obtain an out-of-plane bond order phase (z-BOP) instability below a critical temperature $T_c^{\textrm{z-BOP}}$ which decreases with increasing the doping $\delta$ ending at $T = 0$ at a quantum critical point (QCP) below quarter filling ($\delta=0.5$).
We study the competition between the z-BOP and superconductivity, and find a dome  behavior for superconducting $T_c$ at $\delta < 0.5$. We also discuss the possible role of z-BOP for the dome behavior for  $T_c$ as a function of pressure. 
 
The paper is organized as follows.
In Sec. II we present generic theoretical large-$N$ results as a function of $t'$, $\delta$, $J_\perp$, and $T$. In addition we discuss superconductivity, the z-BOP and their competition. Then, our results can also be considered as a complement of previous different theoretical studies \cite{lu24,qu24}. In Sec. III we discuss a possible and \textit{qualitative} contact with the phenomenology of pressurized nickelates. 
In Sec. IV we present the conclusion and discussions.  

\section{Generic theoretical results}

\subsection{The model and superconductivity due to $J_\perp$}

We study the bilayer $t$-$J_\parallel$-$J_\perp$ model \cite{lu24,qu24},  where only the $d_{x^2-y^2}$ orbitals are considered active
\begin{align}
H =&
\sum_{i, j,\sigma,\alpha} t_{ij} \, \tilde{c}^\dag_{i\sigma,\alpha} \tilde{c}_{j\sigma,\alpha} +
J_\parallel \sum_{i,j,\alpha} \left( \vec{S}_{i,\alpha} \cdot \vec{S}_{j,\alpha} - \frac{1}{4} n_{i,\alpha} n_{j,\alpha} \right)
\nonumber\\
&+ \frac{J_\perp}{2} \sum_{i,\alpha} \left( \vec{S}_{i,\alpha} \cdot \vec{S}_{i,\bar{\alpha}} - \frac{1}{4} n_{i,\alpha} n_{i,\bar{\alpha}} \right) \nonumber \\
&+ t_\perp \sum_{i,\sigma,\alpha} \tilde{c}^\dag_{i\sigma,\alpha}\tilde{c}_{i\sigma,\bar{\alpha}}
- \mu \sum_{i,\alpha} n_{i,\alpha}\,.
\label{eq:H}  
\end{align}
This one orbital bilayer model is derived from the bilayer two orbitals ($d_{x^2-y^2}$ and $d_{z^2}$) model in the limit of strong Hund's coupling $J_H$ \cite{lu24,qu24}. 
Although the interlayer spin exchange is small for the $d_{x^2-y^2}$ orbitals, the large ferromagnetic Hund's coupling can
transfer the interlayer coupling from the $d_{z^2}$ to the $d_{x^2-y^2}$ orbitals. Albeit the validity of the bilayer $t$-$J_\parallel$-$J_\perp$ model for describing the full phenomenology of nickelates is still unknown and the $d_{z^2}$ orbital may play an active role, the model is considered to be one of the prototype models describing high-T$_c$ superconductivity. Hence, it is important to study this model from  a different perspective, because it is possible that interesting characteristics and novel correlated phased can be revealed. The fact that $T_c$ has high values and normal state properties indicate the possible presence of correlations at the high doping $\delta=0.5$, suggesting the important role of $J_\perp$ because, as we will see later in this section, in-plane correlations seem to be unable to trigger high-$T_c$ values at $\delta=0.5$.

In Eq. (\ref{eq:H}) $\alpha$ runs over the planes $1$ and $2$, $\bar{\alpha}$ represents the plane opposite to $\alpha$, and $i$ and $j$ run over the sites of the square lattice of each plane.
The hopping $t_{ij}$ takes a value $t$ between the first nearest-neighbors and $t'$ between the next nearest-neighbors sites on each plane, $\mu$ is the chemical potential.
The exchange interaction $J_\parallel$, 
takes a value between the first nearest-neighbor sites on each plane.
The exchange interaction and the hopping integral 
between planes are $J_\perp$ and $t_\perp$, 
respectively. 
$\tilde{c}^\dag_{i\sigma,\alpha}$ \mbox{($\tilde{c}_{i\sigma,\alpha}$ )} is 
the creation (annihilation) operator of electrons in site $i$ of the plane $\alpha$ with spin $\sigma(=\uparrow, \downarrow)$ in the Fock space without double occupancy.
$n_{i,\alpha}=\sum_{\sigma} \tilde{c}^\dag_{i\sigma,\alpha}\tilde{c}_{i\sigma,\alpha}$ 
is the electron density operator and $\vec{S}_{i,\alpha}$ is the spin operator.
For a given filling $\delta$ we compute
the corresponding chemical potential $\mu$
in a self-consistent manner, as discussed in 
Appendix \ref{app:fullformalism}, which also contains 
a complete description of the path
integral large-$N$ formulation 
of our model~\eqref{eq:H}.

\begin{figure}[t]
\centering
\includegraphics[width=8.5cm]{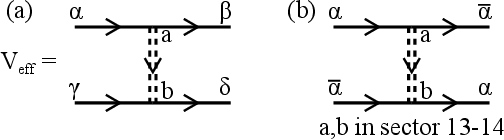}
\caption{(a) General superconducting effective interaction between fermions mediated by the full $14\times 14$ bosonic propagator $D_{ab}$ [Eq. (\ref{eq:Dyson})] (double dashed line).
(b) Out-of-plane superconducting effective interaction mediated by the bosonic propagator corresponding to the $J_\perp$ subsector $r_\perp$-$A_\perp$ [Eq. (\ref{eq:DzBOP})].}
\label{figVeff}
\end{figure}

The general superconducting effective interaction between a fermion with momentum ${\bf k}$ and energy $\nu_n$ and other with momentum ${\bf k}'$ and energy $\nu'_n$,
$V_{{\rm eff}}({\bf k},{\bf k'};\nu_n,\nu'_n)$, can be calculated using the diagram in Fig. \ref{figVeff}(a), which shows that in the present theory pairing is mediated by the $14 \times 14$ bosonic propagator $D_{ab}$ [Eq. (\ref{eq:Dyson})] formed by the $14$-component bosonic field defined as:
\begin{align}
\delta X^{a} =& (\delta R_1,\;\delta{\lambda_1},\;
                \delta R_2,\;\delta{\lambda_2},\;
                r_1^{x},\;r_1^{y},\; A_1^{x},\;A_1^{y},\; \nonumber \\
              & r_2^{x},\;r_2^{y},\; A_2^{x},\;A_2^{y},\;
                r_\perp,\; A_\perp)\, ,
\label{eq:boson-fieldp}
\end{align}
where $\delta R_\alpha$ describes fluctuations of the number of holes at a given site in the plane $\alpha$ and  it is related to on-site charge fluctuations in each plane, $\delta \lambda_\alpha$ is the fluctuation of the Lagrange multiplier introduced to enforce the constraint that prohibits the double occupancy at any site in each plane, and $r_\alpha^{x}$ and $r_\alpha^{y}$ ($A_\alpha^{x}$ and $A_\alpha^{y}$) describe fluctuations of the real (imaginary) part of the in-plane bond field coming from the $J_\parallel$-term.
$r_\perp$ and $A_\perp$ are the real and imaginary part, respectively,  of the out-of-plane bond field coming from the $J_\perp$-term.
See Appendix \ref{app:fullformalism} for technical details.

The analytical expression for the effective interaction is
\begin{equation}
V_{{\rm eff}}({\bf k},{\bf k'};\nu_n,\nu_n') = \Lambda_{\alpha\beta,a} D_{ab}({\bf k}-{\bf k'},\nu_n-\nu_n') \Lambda_{\gamma\delta,b} \, ,
\label{eq:veff}
\end{equation}
\noindent where $\Lambda_{\alpha\beta,a}$ and $\Lambda_{\gamma\delta,b}$ are three-legs vertices [Eqs. (\ref{eq:threeleg1}) and (\ref{eq:threeleg2})] describing the interactions between fermions and bosons, where its  momentum and frequency dependence are omitted for simplicity.
The indices $\alpha$, $\beta$, $\gamma$, and $\delta$ indicate the plane index (1 or 2), and the indices $a$ and $b$ the boson flavor [Eq. (\ref{eq:boson-fieldp})].
Note that we can also draw a diagram containing two four-legs vertices $\Lambda_{\alpha\beta,ab}$ [Eq. (\ref{eq:fourleg})] and two bosonic propagators $D_{ab}$, however, this contribution is omitted because it is $O(1/N^2)$.

In Appendix \ref{app:jperp} we focus on the $J_\perp$-subspace of the full formulation. This is mainly motivated by the fact that in first approximation, the $J_\perp$-sector 
of the $14 \times 14$ matrix $D_{ab}$ [Eq. (\ref{eq:Dyson})] contains $J_\perp$, which might be a relevant quantity for superconductivity at $\delta=0.5$. 
In Fig. \ref{figVeff}(b) we show the diagram for the out-of-plane effective interactions corresponding to the exchange term $J_\perp$ of Eq. (\ref{eq:H}), i.e., sector 13-14 (also called $r_\perp$-$A_\perp$ sector) of Eq. (\ref{eq:boson-fieldp}).
Using the bare bosonic propagator [Eq. (\ref{eq:Jperpsec})], which is equivalent to a mean-field calculation \cite{zinni21}, and the three-legs vertices [Eq. (\ref{eq:Jperpvert})], it is straightforward to show that the effective superconducting interaction from  Fig. \ref{figVeff}(b) reads as
\be
V^{\rm{eff}}_\perp(\vk,\vk')= \frac{J_\perp}{4}, 
\ee
\noindent which is clearly isotropic $s$-wave and unretarded.
This result is consistent with the slave-boson mean-field approach \cite{lu24} where the pairing strength is $\sim J_\perp$. 

Similarly, the main in-plane contribution to superconductivity is given by the $d$-wave unretarded effective interaction \cite{zinni21} 
\be
V^{\rm{eff}}_\parallel(\vk,\vk')= J_\parallel \gamma_d(\vk) \gamma_d(\vk'),
\label{eq:Veffpar}
\ee
\noindent with $\gamma_d({\bf k}) = \frac{1}{2}(\cos k_x - \cos k_y)$.
This result can also be obtained in the present approach (Appendix \ref{app:fullformalism}) if we focus on the in-plane $J_\parallel$-sector, i.e., the sector 1-12 of [Eq. (\ref{eq:boson-fieldp})]. 
We note that the
presence of unretarded effective interactions, also known as pairing without glue, in cuprates is still under discussion \cite{anderson07,maier08,yamase23,zinni21}.

\begin{figure}[t]
\centering
\includegraphics[width=8.6cm]{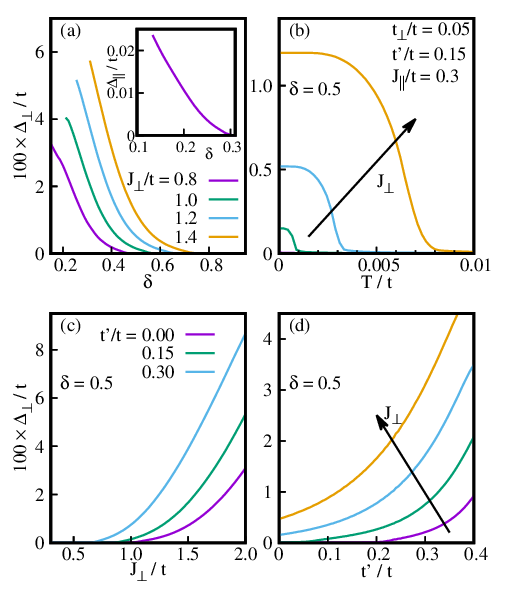}
\caption{Superconductivity in the bilayer $t$-$J_\parallel$-$J_\perp$ model. (a) Out-of-plane superconducting gap $\Delta_\perp$ versus $\delta$ for $J_\perp/t=0.8,1.0,1.2,1.4$.
(b) $\Delta_\perp$ versus $T$ for $J_\perp/t=1.0,1.2,1.4$ at $\delta=0.50$. (c) $\Delta_\perp$ versus $J_\perp$ for $t'/t=0, 0.15, 0.3$ at $\delta=0.5$.
(d) $\Delta_\perp$ versus $t'$ for $J_\perp/t=0.8,1.0,1.2,1.4$ at $\delta=0.5$.
The inset in (a) shows the in-plane superconducting gap $\Delta_\parallel$ versus $\delta$ for $J_\parallel/t=0.3$ and $J_\perp=0$ at $T=0$. }
\label{fig2_ni}
\end{figure}

The presence of $V^{\rm{eff}}_\parallel(\vk,\vk')$  and  $V^{\rm{eff}}_\perp(\vk,\vk')$ indicates the possibility of in- and out-of-plane superconductivity, respectively, and their interaction.
In Appendix \ref{app:scgapeq} we present the corresponding 
two-gap formulation, where the equations for the superconducting in-plane $\Delta_\parallel$ and out-of-plane $\Delta_\perp$ gaps, Eq. (\ref{eq:gappar}) and Eq. (\ref{eq:gapper}) respectively, are coupled to each other. In this paper, we present results for $J_\parallel/t = 0.3$ similar to cuprates \cite{keimer15} and $t_\perp/t = 0.05$ \cite{lu24}, while we vary $t'$, $J_\perp$, $\delta$ and the temperature $T$.

In Fig.~\ref{fig2_ni}(a) we plot the out-of-plane superconducting gap $\Delta_\perp$ versus doping for $J_\perp/t=0.8, 1.0, 1.2$, and $1.4$, at $T=0$.
We obtain an out-of-plane $s$-wave superconductivity triggered by $J_\perp$ for all dopings. Note that in spite of $J_\parallel/t = 0.3$, in Fig.~\ref {fig2_ni}(a) the in-plane $\Delta_\parallel = 0$ for all dopings, i.e., $J_\perp$ disfavors in-plane $d$-wave  superconductivity, as was also discussed in \cite{oh24t}. 
For completeness, in the inset of Fig.~\ref{fig2_ni}(a) we show results for $J_\perp=0$.
Since $J_\parallel/t = 0.3$  is a condition similar to cuprates we obtain in-plane $d$-wave superconductivity.
In fact, we show an in-plane $\Delta_\parallel$ gap versus doping which vanishes for doping $\delta \gtrsim 0.3$ as observed in cuprates \cite{keimer15}.
Thus, it is clear that we cannot expect a large value for $T_c$ at quarter filling ($\delta = 0.5$), unless we use unrealistic values for $J_\parallel/t \sim 2$.
Besides, this value for $J_\parallel$ is unrealistic, it triggers several instabilities,   including phase separation, that may cover mainly the full phase diagram \cite{bejas12}.  
The fact that the in-plane $d$-wave superconductivity triggered by $J_\parallel$ is far from the doping of interest for pressurized nickelates justifies that in first approximation   $V^{\rm{eff}}_\parallel$ from the  
$J_\parallel$-sector can be studied separately from the $J_\perp$-sector. 
In Fig.~\ref{fig2_ni}(b), for $\delta=0.5$, we plot $\Delta_\perp$ for $J_\perp=1.0$, $1.2$, and $1.4$ versus temperature, and obtain the  superconducting critical temperatures $T_c/t \sim 0.001$, $0.003$, and $0.007$, respectively. In agreement with Fig.~\ref{fig2_ni}(a) there is no superconductivity for $J_\perp/t=0.8$ at quarter filling. 

In Fig.~\ref{fig2_ni}(c) we show $\Delta_\perp$ versus $J_\perp$ for $\delta = 0.5$, and $t'/t=0, 0.15$, and $0.3$.
The figure shows that the superconducting gap at quarter filling, and the corresponding $T_c$,  increases with increasing $J_\perp$ and it is only expected for $J_\perp/t$ bigger than a given value that decrease with increasing $t'$.
The in-plane $d$-wave superconducting gap is $\Delta_\parallel = 0$ for all $J_\perp$ in spite of $J_\parallel/t = 0.3$, which is due to the large doping value $\delta = 0.5$. A reasonable variation of $t_\perp$ will not effectively affect $\Delta_\perp$. However, a large $t_\perp$, comparable to $J_\perp$, will suppress $\Delta_\perp$ due to the Pauli blocking effect \cite{qu24}.
In Fig.~\ref{fig2_ni}(d) we show that $\Delta_\perp$ also increases with increasing the in-plane $t'$ for all values of $J_\perp$ chosen. In summary, tendencies to out-of-plane $s$-wave superconductivity increase with increasing $J_\perp$ and $t'$, and in addition there is a lower limit for $J_\perp$ below which  superconductivity can not be expected. 

\subsection{Out-of-plane bond order phase z-BOP}
\label{sec:zBOP}

In this section, we investigate the instability in the charge channel. In particular, we focus on an out-of-plane bond order (z-BOP). For discussing the z-BOP we focus on the $r_\perp$-$A_\perp$ sector of the matrix $D_{ab}$, i.e., the dressed $D_{\textrm{z-BOP}}$ [Eq. (\ref{eq:DzBOP})] which is beyond mean field level. 
When for a given doping, temperature $T_c^{\textrm{z-BOP}}$, and momentum ${\bf q}_c$, one eigenvalue of $D_{\textrm{z-BOP}}^{-1}$ is zero, the out-of-plane bond order instability takes place.
The order parameter of the z-BOP is defined in Eq. (\ref{eq:phi4x4}). 
This order parameter corresponds to a charge instability and is associated with the interplane bond order $\langle f_{{\bf k} \sigma,1}^{\dagger} f_{{\bf k}+{\bf q}_c \sigma,2} \rangle$, where the operator $f$ is associated to the Hubbard operator $\tilde{c}$ (see Appendixes A and D for more detail). Thus, the freezing of $r_\perp$ or $A_\perp$ at the transition indicates a modulation of the interlayer hopping with wave vector ${\bf q}_c$.

\begin{figure}[t]
\centering
\includegraphics[width=8.6cm]{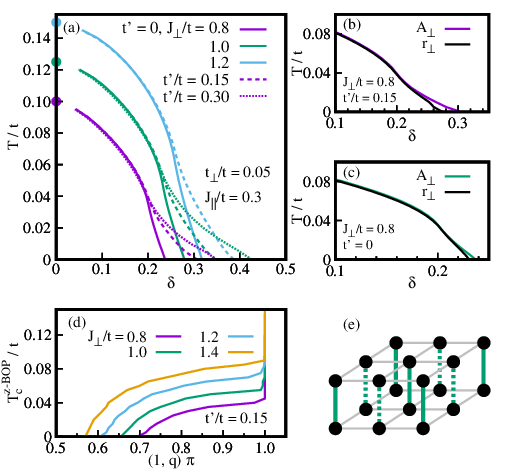}
\caption{(a) $T_c^{\textrm{z-BOP}}$ versus doping for different values of $J_\perp$ and $t'$. The points on the $y$-axis are located at $J_\perp/8$ where each $T_c^{\textrm{z-BOP}}$ extrapolates to half-filling ($\delta=0$).
(b) $T_c^{\textrm{z-BOP}}$, for both sectors $r_\perp$ and $A_\perp$, for $J_\perp/t=0.8$ and $t'/t=0.15$.
(c) The same as (b) but for $t'=0$. (d) Values of ${\bf q}_c$ and $T_c^{\textrm{z-BOP}}$ along the instability lines for various hole doping with $t'/t=0.15$ and $J_\perp/t=0.8, 1.0, 1.2, 1.4$. 
(e) Sketch of the z-BOP where the out-of-plane bonds are frozen following a pattern with ordered momentum ${\bf q}_c=(\pi,\pi)$. The solid and dashed lines indicate positive and negative renormalization of the out-of-plane hopping, respectively.}
\label{fig3pp_ni}
\end{figure}

In Fig.~\ref{fig3pp_ni} (a) we show $T_c^{\textrm{z-BOP}}$ versus doping for different values of $J_\perp$ and $t'$.
Clearly, with increasing $J_\perp$ and $t'$ the z-BOP expands its area in the phase diagram.
Instead, z-BOP is nearly independent of $t_\perp$ (not shown).
In all cases, $T_c^{\textrm{z-BOP}}$ decreases with increasing doping ending at a QCP at $T = 0$.
The critical momentum ${\bf q}_c$ is incommensurate and of the form ${\bf q}_c=(1, q)\pi$, with $q \sim 0.55$-$0.70$ at $T = 0$, and moves towards ${\bf q}_c = (\pi,\pi)$ with increasing $T$ [see Fig.~\ref{fig3pp_ni}(d)], where we have plotted ${\bf q}_c$ along $T_c^{\textrm{z-BOP}}$, i.e., along the instability lines for various hole doping in Fig.~\ref{fig3pp_ni}(a).
Thus, below  $T_c^{\textrm{z-BOP}}$ the z-BOP is the stable phase. 

We note that for the present parameters the z-BOP occurs at much larger doping than the flux phase \cite{bejas12,affleck88a,cappelluti99,morse91}, or $d$CDW \cite{chakravarty01,Tu16,Choubey17,Chakraborty19}, in the two-dimensional (2D) $t$-$J$ model.
For instance, for $J_\parallel/t = 0.3$, and $t'=0$, while for $J_\perp = 0$ the flux phase occurs at $\delta \lesssim 0.13$ at $T = 0$ \cite{bejas12,morse91}, the z-BOP for $J_\perp/t = 0.8$ occurs at
$\delta \lesssim 0.24$.
This fact justifies, in first approximation,  the study of the z-BOP from the  $J_\perp$-sector separately because it is relevant in  the doping regime of our main interest, i.e., $\delta \sim 0.5$. 

However, we have to take care close to half-filling and low temperature. While 
at $\delta = 0$, $T_c^{\textrm{z-BOP}} = J_\perp/8$ [see the points on the $y$-axis in  Fig.~\ref{fig3pp_ni}(a)], the bond order instabilities, including the flux phase, in the pure 2D $t$-$J$ model develops at $T=J_\parallel/8$ \cite{bejas12}. Then, at low doping and low $T$ a competition between the 2D bond order instabilities and z-BOP is expected. Such analysis requires the full approach (Appendix \ref{app:fullformalism}), which is a big challenge that demands the stability analysis of the $14\times 14$ $D_{ab}$ matrix and deserves a separate theoretical study.

At $\delta=0$ one would expect no charge dynamics in the usual $t$-$J$ model. However, in the large-$N$ limit, the charge carriers have an additional effective in-plane (out-of-plane) hopping given by $\chi$ [Eq. (\ref{chi})] ($\chi'$ [Eq. (\ref{chi'})]) triggered by $J_\parallel$ ($J_\perp$).
Even though the validity of the large-$N$ expansion can be debatable at half-filling or at very low doping close to half-filling, it has been useful for describing charge excitations at finite doping \cite{greco16,nag20,hepting22,hepting23a,nag24}. Hence, we consider that the method can be faithfully applied near $\delta=0.5$ where superconductivity is observed in nickelates.

The leading instability of the z-BOP occurs in the sector $A_\perp$, i.e., the associated eigenvector of $D_{\textrm{z-BOP}}^{-1}$ [Eq. (\ref{eq:DzBOP})] is of the form $(0,1)$.
However, the instability in the sector $r_\perp$ is very close and even both merge at high temperature, as can be seen in Fig.~\ref{fig3pp_ni}(b) and (c).
Thus, the instabilities in the $r_\perp$ and $A_\perp$ sectors are nearly degenerated.
In addition, with decreasing $t'$ the instabilities in both channels are even
closer at $T=0$, as can be seen by comparing results for $J_\perp/t=0.8$ for $t'/t=0.15$   [Fig.~\ref{fig3pp_ni}(b)] and $t'=0$ [Fig.~\ref{fig3pp_ni}(c)].

The projection of $D_{\textrm{z-BOP}}$ on the two eigenvectors $(1, \, 0)$ and $(0, \, 1)$ leads to the z-BOP susceptibilities $\chi^{r_\perp}_{\textrm{z-BOP}}$ and $\chi^{A_\perp}_{\textrm{z-BOP}}$, Eq. (\ref{eq:sus13}) and Eq. (\ref{eq:sus14}), respectively. Thus, beyond mean-field level the out-of-plane bond susceptibilities diverge at a temperature $T_c^{\textrm{z-BOP}}$ and momentum ${\bf q}_c$.

\begin{figure}[t]
\centering
\includegraphics[width=8.6cm]{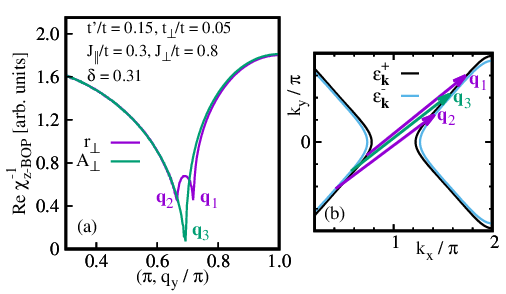}
\caption{(a) The inverse of the static z-BOP susceptibilities $\chi^{r_\perp}_{\textrm{z-BOP}}$ and $\chi^{A_\perp}_{\textrm{z-BOP}}$ along the $(\pi,0)$-$(\pi,\pi)$ direction.
(b) ${\bf q}_1$ and ${\bf q}_2$ are the two intra-band nesting wave-vectors associated with the two peaks of $\chi^{r_\perp}_{\textrm{z-BOP}}$, while ${\bf q}_3$ corresponds to interband nesting wave-vector associated with the single peak in $\chi^{A_\perp}_{\textrm{z-BOP}}$.}
\label{2kF}
\end{figure}

In Fig.~\ref{2kF}(a) we show the real part of the inverse of the static [$\chi^{r_\perp,A_\perp}_{\textrm{z-BOP}}({\bf q}, \omega=0)$] z-BOP susceptibilities for $J_\perp/t=0.8$ at $\delta \sim 0.31$ close to the QCP [Fig.~\ref{fig3pp_ni}(b)], and  momentum ${\bf q}$ along the $(\pi,0)$-$(\pi,\pi)$ direction where the instability takes place.
As it can be seen, the susceptibilities show strong peaks close to the incommensurate momentum ${\bf q}_c \sim (\pi,0.7\pi)$ [see also Fig.~\ref{fig3pp_ni}(d)]. When these peaks reach the zero value the instability occurs. We can further ask: what is the origin of the z-BOP instability? The origin lies on the existence of nesting wave-vectors. As shown in Fig.~\ref{2kF}(b), while the instability in the $r_\perp$ sector is related to two intra-band nesting wave-vectors ${\bf q}_1$ and ${\bf q}_2$, the instability in the $A_\perp$ sector is related to an interband nesting wave-vector ${\bf q}_3$.

 Since the instabilities $\chi^{r_\perp}_{\textrm{z-BOP}}$ and $\chi^{A_\perp}_{\textrm{z-BOP}}$ are nearly degenerated any small change in the model can push the instability in one channel in favor of the other, or even one can expect a mixing of both.
 While the instability in the sector $\chi^{r_\perp}_{\textrm{z-BOP}}$ indicates the freezing of the real part of the out-of-plane bond field $r_\perp$, the instability in the sector $\chi^{A_\perp}_{\textrm{z-BOP}}$ indicates the freezing of the imaginary part of the out-of-plane bond field $A_\perp$. In addition, the z-BOP follows a lattice pattern with ordered momentum ${\bf q}_c$. It is important to emphasize that ${\bf q}_c=(\pi,\pi)$ occurs for intermediate doping $0.2 < \delta < 0.3$ and temperature $T \sim 0.04$ [Fig.~\ref{fig3pp_ni}(a)]. On the other hand Fig.~\ref{fig3pp_ni}(d) shows that ${\bf q}_c$ is, in general, close to $(\pi,\pi)$.
In Fig.~\ref{fig3pp_ni}(e), for simplicity,  we sketched a picture of such pattern in the real space for the commensurate case ${\bf q}_c=(\pi,\pi)$.
Solid (dashed) lines in Fig.~\ref{fig3pp_ni}(e) indicate that the out-of-plane hopping $(t_\perp \delta/2 - \chi')$ is renormalized with a positive (negative) contribution proportional to the z-BOP gap $\phi$ (see Appendix \ref{app:zBOPsus} for more details).

In addition, the presence of a z-BOP below $T_c^{\textrm{z-BOP}}$ opens a gap $\phi$ in the system. In Appendix \ref{app:zBOPsus} we present the gap equation for  $\phi$, which for simplicity is considered to be real, and we assumed the incommensurate wave vector ${\bf q}_c$ for each $J_\perp$ [Fig.~\ref{fig3pp_ni}(d)] as the ordered momentum of the z-BOP. Then, we focus on the instability in the real sector from $\chi^{r_\perp}_{\textrm{z-BOP}}$, which on the other hand is very close to the instability from the imaginary sector given by $\chi^{A_\perp}_{\textrm{z-BOP}}$. Thus, the present studied z-BOP is a kind of interlayer charge density wave triggered by $J_\perp$.

\begin{figure}[t]
\centering
\includegraphics[width=8.6cm]{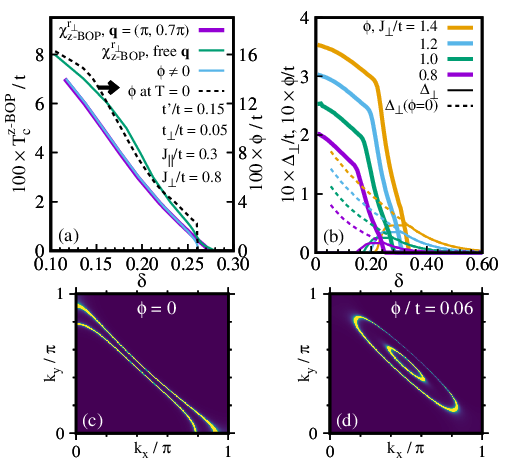}
\caption{(a) Green solid line shows $T_c^{\textrm{z-BOP}}$ versus doping. Violet solid line is the same as green solid line but searching for the instability for a fix ${\bf q} = (\pi, 0.7\pi)$. Blue solid line shows $T_c^{\textrm{z-BOP}}$ versus doping obtained when the z-BOP gap $\phi$ becomes nonzero. Black dashed line shows  $\phi$ (measured on the right $y$-axis scale) versus doping at $T=0$. (b) The superconducting $\Delta_\perp$ (thin lines) and z-BOP $\phi$ gaps (thick lines) versus doping for the competing case (solid lines) for $J_\perp/t=0.8$, $1.0$, $1.2$, and $1.4$. Dashed lines show the superconducting $\Delta_\perp$ versus doping without competition with the z-BOP, same as Fig. \ref{fig2_ni}(a).
(c) The two-sheets FS for $\delta=0.20$ and $\phi=0$. (d) The same as (c) for $\phi/t = 0.06$ where, for simplicity, an ordered momentum ${\bf q}_c=(\pi,\pi)$ was chosen.
}
\label{just}
\end{figure}

Now, we compare the solution of the gap equation for $\phi$ with the instability temperatures obtained from the susceptibilities. In Fig.~\ref{just}(a) we present results that show that the  gap equation for $\phi$  represents a reasonable approximation.
Solid green line is the result showed in Fig.~\ref{fig3pp_ni}(b) for $T_c^{\textrm{z-BOP}}$ versus doping for $J_\perp/t=0.8$ and $t'/t=0.15$.
Violet line is $T_c^{\textrm{z-BOP}}$ obtained from the instability of $\chi^{r_\perp}_{\textrm{z-BOP}}$ for a fix ${\bf q}_c = (\pi,0.7\pi)$.
The difference between green and violet lines occurs at temperatures where the ordered momentum moves from $(\pi,0.7\pi)$ towards $(\pi,\pi)$ [see Fig. \ref{fig3pp_ni}(d)].
Blue solid line shows $T_c^{\textrm{z-BOP}}$ versus doping obtained when the z-BOP gap $\phi$ becomes nonzero. Thus, $T_c^{\textrm{z-BOP}}$ is the temperature where the z-BOP susceptibility diverges and, as should be, the z-BOP gap becomes nonzero. Black dashed line shows $\phi$ versus doping at $T=0$.  $\phi$ decreases with increasing doping and goes to zero at a QCP [Fig.~\ref{fig3pp_ni}(b)].

In Fig.~\ref{just}(c) we show the two-sheets Fermi surface (FS) for $\delta=0.20$ in the normal state ($T > T_c^{\textrm{z-BOP}}$, i.e., $\phi=0$). The two-sheets splitting comes from $t_\perp$ and the contribution $\chi'$ [see Eq. (\ref{eq:G0}) and Eq. (\ref{chi'})] to the interplane hopping.
Figure \ref{just}(d) shows the two-sheets FS surface at $T=0$ for $\phi/t=0.06$. Here, for simplicity we use ${\bf q}_c=(\pi,\pi)$ instead of ${\bf q}_c=(\pi,0.7\pi)$.
Thus, inside the z-BOP, below $T_c^{\textrm{z-BOP}}$, the FS is formed by split pockets, indicating that in the z-BOP the FS is not fully gapped, and the system remains metallic with a possible pseudogap-like feature.

\subsection{Competition between z-BOP and superconductivity}

Having showed the results for the superconductivity and z-BOP separately, in the last two sections, we now look into the competition between the two. According to Fig.~\ref{fig2_ni}(a) $T_c$  increases with decreasing doping. On the other hand, in the above section we described the existence of a z-BOP instability triggered by $J_\perp$. Then, 
as in the case of the $d$CDW in the 2D $t$-$J$ model \cite{zeyher03}, 
a competition between both phases, i.e., superconductivity and z-BOP, is expected. 

For discussing the above picture, in Appendix \ref{app:competition} we propose a  theoretical framework for studying the superconducting $\Delta_\perp$, the z-BOP gap $\phi$, and their competition [Eq. (\ref{eq:Deltaperp}) and Eq. (\ref{eq:phi})], where we consider the incommensurate wave vector ${\bf q}_c$ as the ordered momentum of the z-BOP.

Figure \ref{just}(b) shows results at $T=0$ for $J_\perp/t=0.8$, $1.0$, $1.2$, and $1.4$, where we use the corresponding ${\bf q}_c$ for each $J_\perp$ [Fig.~\ref{fig3pp_ni} (d)].  
The superconducting gaps $\Delta_\perp$ for each case show a dome behavior with an optimal doping at $\delta \sim 0.25$-$0.35$ close to the doping where the z-BOP is developed for each $J_\perp$. Without competition between superconductivity and z-BOP, i.e., if $\phi=0$ is assumed in the calculation,  $\Delta_\perp$ continues increasing with decreasing doping as showed by the dashed lines. In other words, if z-BOP is removed superconductivity tends to increase with decreasing doping, as we mentioned when discussing Fig. \ref{fig2_ni}(a).

A recent density matrix renormalization group study in the type II $t$-$J$ model for bilayer nickelates \cite{oh24told} also suggests the existence of a superconducting dome with a maximum at $\delta \sim 0.4$-$0.5$. 

\section{Possible implications with the phenomenology of pressurized nickelates.}
\label{sec_III}

In this section we discuss a \textit{possible} contact  between our results and pressurized nickelates. Considering the fact that our results are obtained on the basis of a \textit{simplified} model in a \textit{given approximation}, our discussions should be considered \textit{qualitatively}. 

Since the value and the pressure dependence of $J_\perp$ is unknown, in Sec. II we have analyzed our results as a function of $J_\perp$ as in previous references \cite{qu24,lu24}. Some reports suggest that $J_\perp \sim 2J_\parallel \sim 460$ meV \cite{lu23}, and $J_\perp=403$ meV
\cite{qu23p},
$J_\perp=320$ meV \cite{qu24}, $J_\perp \sim t$, i.e., $0.5$eV or even larger \cite{oh2024}.
 
Just as an example of an estimation, for $\delta = 0.50$, $J_\perp/t=1.4$, and $t'/t=0.15$ \cite{luo23} we obtain a superconducting critical temperature $T_c/t \sim 0.007$ [Fig.~\ref{fig2_ni}(b)], which is close to $T_c \sim 80$ K, i.e., of the same order as that reported in bilayer nickelates \cite{sun23}. 
For this estimate we have assumed $t/2 = 0.5$~eV \cite{luo23,misc-factor2a}. We call attention that $J_\perp/t=1.4$ is somewhat large, but of the order of some estimations \cite{oh2024}. Maybe other facts can also play a role for quantitative comparisons. For instance, in our large-$N$ approximation the prefactor in the gap equation Eq. (\ref{eq:gapper}) is $J_\perp/4$, while at the level of slave boson mean field theory \cite{lu23} it is $3J_\perp/8$ which may be favorable for larger $T_c$ values for a smaller $J_\perp$ than the large-$N$.

If we assume  that the increase of pressure induces an increase of $J_\perp$, the result of 
Fig.~\ref{fig2_ni}(c) offers a \textit{possible} explanation for the observation that $T_c$ increases with pressure. In addition,  superconductivity is expected only if $J_\perp$ is larger than a certain value which, for instance, is $J_\perp/t \sim 0.9$ for $t'/t=0.15$ at $\delta=0.5$. This feature might explain the presence of a critical pressure (i.e., a critical $J_\perp$) for the appearance of superconductivity, as observed in experiments.

Interestingly, if $J_\perp$ is vanishingly small
only in-plane $d$-wave superconductivity is possible below $\delta \lesssim 0.30$ [inset of Fig.~\ref{fig2_ni}(a)], similar to cuprates where $J_\perp=0$ or negligible.
It can be shown that for $\delta = 0.16$, which corresponds approximately to optimally doping, the superconducting critical temperature is about $T_c/t \sim 0.007$, i.e.,  $T_c \sim 80$ K which is similar to that observed in cuprates \cite{keimer15}. However, if $J_\perp > J_\parallel$, as discussed for nickelates \cite{lu24}, our theory predicts that $J_\perp$ disfavors in-plane $d$-wave superconductivity for all dopings [see Fig.~\ref{fig2_ni}(a)]. 

To our knowledge there are no reports yet of underdoped and/or overdoped nickelates under pressure. But certainly, it is expected and desirable that these experiments come in the near future, 
and they are important for the understanding of the phenomenology of these materials. If at quarter filling (i.e., $\delta=0.5$) we already have a large value of $T_c$ we can expect larger values for $T_c$ at lower doping. In fact, this is a key feature in the 2D $t$-$J$ model in the context of cuprates. The lower the doping is, the less diluted are the charge carriers and $J$ is more effective for pairing.
According to Fig.~\ref{fig2_ni}(a) $T_c$  increases monotonically with decreasing doping,  reaching possibly  unrealistic high values of $T_c$. \textit{One  possible expectation}  is that at lower dopings some instability to a new phase develops, competes with superconductivity, and pushes down $T_c$ as in cuprates. Beyond mean-field level we have obtained a z-BOP which can compete with superconductivity, leading to a dome in $T_c$ as 
a function of doping, as shown in Fig.~\ref{just}(b).  The dome presented in Fig.~\ref{just}(b) has optimal dopings at $ \delta \sim 0.25$-$0.35$ depending on the value of $J_\perp$.
Of course this analysis must be taken with caution because it assumes that the system can be doped at a fixed pressure keeping $J_\perp$ rather constant, which is in  fact unknown and nontrivial. A recent report \cite{meng2024} suggests the presence of two distinct CDWs, one at ambient pressure which tends to disappear with increasing pressure, and the other that develops at high pressure. Although it is too early for affirmations we just have in mind that z-BOP is triggered by $J_\perp$ which may increase with increasing pressure.

Since the z-BOP is basically a charge instability it would be interesting to search for that using x-ray techniques,
as those used for studying the charge order in cuprate superconductors \cite{keimer15}. On the other hand, z-BOP and/or its fluctuations can get coupled with the lattice and phonons. 
In addition, the z-BOP leads to a reconstruction of the FS [Fig.~\ref{just}(d)] and, in principle, can be detected by ARPES.
From the theory side, it would be also interesting to see if the z-BOP is obtained by using other theoretical methods. For instance, we believe that the slave-boson approach  beyond mean-field would have to show the z-BOP, as it shows different bond order phases triggered by $J_\parallel$ in the 2D $t$-$J$ model \cite{morse91}.

\begin{figure}[t]
\centering
\includegraphics[width=8.6cm]{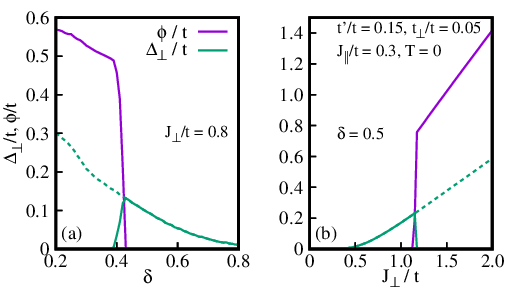}
\caption{(a) The superconducting $\Delta_\perp$ and z-BOP $\phi$ gaps versus doping for the competing case (solid lines). The dashed line shows the superconducting $\Delta_\perp$ versus doping without competition with the z-BOP. (b) $\Delta_\perp$ and $\phi$ as a function of $J_\perp$. Dashed line is the result for $\Delta_\perp$  without the coupling between superconductivity and z-BOP, while for solid lines both phases compete. See text for discussions.}
\label{Comp}
\end{figure}

Given the qualitative aspect of our results, in the following and just as an example, we choose the prefactor $3J_\perp/4$ and $3J_\perp/2$ on the right-hand side of equations Eq. (\ref{eq:Deltaperp}) and Eq. (\ref{eq:phi}), respectively, instead of $J_\perp/4$ and $J_\perp/2$. 
In Fig.~\ref{Comp}(a) we show the results for $J_\perp/t=0.8$; a value closer to the one reported in Ref. \cite{qu24}.
In this case we obtain a dome in $T_c$ with maximum at $\delta \sim 0.45$, closer to $\delta=0.5$. Note also that in a small doping region below $\delta \sim 0.45$, superconductivity and z-BOP coexists.

Some reports in pressurized nickelates discuss the existence of a strange metal phase at high temperatures \cite{sun23,craco24,zhang23}.
Then, it will be very interesting to perform experiments as a function of doping and see if a dome in $T_c$ is observed, which may indicate the presence of a pseudogap, a strange metal phase at high temperatures around optimal doping as in cuprates, 
and in addition a Fermi liquid phase at low temperatures and high doping.
Whether a possible pseudogap can be associated with z-BOP is not straightforward, but it is suggestive that, as in the 2D $t$-$J$ model \cite{zeyher03}, a bond-order instability might compete with superconductivity and leads to a dome behavior for $T_c$.

In Ref. \cite{li24pppp}  
a dome of  $T_c$ as a function of pressure
was reported, i.e.,
$T_c$ increases with pressure reaching the maximum at 18 GPa, and for larger pressures $T_c$ is suppressed.
If the increasing of pressure means \textit{crudely} an  increase of $J_\perp$, then Fig.~\ref{Comp}(b) could offer a possible explanation: Superconductivity increases until $J_\perp/t \sim 1.1$ where the z-BOP develops, leading to a decrease of $\Delta_\perp$ for larger $J_\perp$ (pressure). Although in the experiment the decrease of T$_{c}$ with pressure is smoother, our approximation shows the main picture.

\section{Discussions and conclusion}
Inspired by the high value of the superconducting critical temperature $T_c$ \cite{sun23} in pressurized nickelates and the similarities with high-$T_c$ cuprate superconductors \cite{hepting23}, we assumed that pressurized bilayer nickelates are strongly correlated systems, and studied the consequences of such assumption.
Since recent experiments indicate that pressurized nickelates are really bulk high-$T_c$ superconductors, a kind of $t$-$J$ model can be expected to describe the physics of these materials. 
In this context, and although its limitations are mentioned in the text, the  $t$-$J_\parallel$-$J_\perp$ model is expected to capture some features of interest. We studied this model in  a large-$N$ approximation based on the path integral representation for Hubbard operators. 
The model contains tendencies to in-plane $d$-wave superconductivity triggered by $J_\parallel$, out-of-plane $s$-wave superconductivity due to $J_\perp$, and the competition of both.
Since superconductivity in nickelates occurs at quarter filling, which is a high hole doping value ($\delta = 0.5$) for the 2D $t$-$J$ model for cuprates, the presence of $J_\perp$ becomes relevant. 

It is well known that the large-$N$ approximation overestimates charge over magnetic excitations \cite{bejas12}. 
Since the $d_{x^2-y^2}$ orbital in nickelates is quarter filling, this large hole doping ($\delta = 0.5$) and the fact that it seems that pressurized nickelates are far from a magnetic order \cite{xie24,chen24pp} 
suggest that the large-$N$ approach might be a good approximation for describing the pressurized nickelates.

$J_\perp$ favors $s$-wave out-of-plane superconductivity at $\delta = 0.5$ [Figs.~\ref{fig2_ni}(a)-\ref{fig2_ni}(b)].
In addition, we found that a lower limit for $J_\perp$ exists in order to get superconductivity [Fig.~\ref{fig2_ni}(c)]. If $J_\perp$ is smaller than this lower limit, superconductivity is not expected at quarter filling,  which can be associated to the fact that the increasing of pressure increases $J_\perp$ and superconductivity emerges. 

At mean-field level we obtain that $T_c$ increases steeply with decreasing doping, reaching possible unrealistic values for $T_c$ at low doping.
However, beyond mean-field level we obtained an instability to an out-of-plane bond-order phase z-BOP below a critical temperature $T_c^{\textrm{z-BOP}}$ with an incommensurate ordered momentum ${\bf q}_c=(1,q)\pi$ at $T=0$.
$T_c^{\textrm{z-BOP}} = 0$ at a QCP close to $\delta \sim 0.25$-$0.40$, depending on the values of $J_\perp$ and $t'$ [Fig.~\ref{fig3pp_ni}(a)].
The competition of z-BOP (and also possibly with its fluctuations) and superconductivity leads to a dome behavior of $T_c$ [Figs.~\ref{just}(b) and \ref{Comp}(a)].
This suggests the possible presence of a pseudogap in pressurized nickelates, similar to cuprates.
This fact may be consistent with the strange metal behavior at high temperature discussed in some reports \cite{sun23,craco24,zhang23}.
Then, it is worthwhile to study the behavior of $T_c$ as a function of doping.

If the increase of pressure increases $J_\perp$, it may explain \textit{ qualitatively} the dome shape behavior for superconductivity as a function of pressure observed experimentally \cite{li24pppp}. With increasing pressure, superconductivity develops at a certain value of pressure, where the  minimum value of $J_\perp$ for triggering superconductivity  is reached. With increasing pressure even more, 
$T_c$ increases until a value of pressure ($J_\perp$) where the z-BOP develops and $T_c$ goes down with a further pressure increase [Fig.~\ref{Comp}(b)]. 

Finally, it is important to mention the very recent discovery of superconductivity in thin films of nickelates at ambient pressure\cite{ko_nat,zhou2024}, which may lead to a big advance by allowing experiments to be performed at ambient pressure, and hence clarifying many open aspects of the topic, like the role of correlations, the symmetry of the superconducting gap, and the basic model which can capture the main physics of these systems.

Before ending, we like to emphasize once more that our results should be considered at a qualitative level and, in spite of the limitations of our approach, we have obtained some positive results arising from this first view of the problem.
This motivates future calculations in the framework of the full approach (Appendix \ref{app:fullformalism}), which of course is a big challenge.
One important and nontrivial issue is about the role of the out-of-plane Coulomb interaction $V_\perp$.
If the increase of pressure increases $J_\perp$, an increase of $V_\perp$ is also expected, as suggested in the context of organic materials \cite{mckenzie97}, and superconductivity can be negligible at mean-field level if $V_\perp \sim J_\perp$. Preliminary calculations beyond mean-field suggest, as in 
Ref.\cite{zinni21} for the 2D $t$-$J$ model, that 
the main role of the dressed $D_{ab}$ [Eq. (\ref{eq:Dyson})] on superconductivity  is to screen out $V_\perp$ due to the fluctuations of $\delta \lambda_{\alpha}$ [Eq. (\ref{eq:boson-field})] which enforce the nondouble occupancy constraint. 

\acknowledgments
The authors thank M. Hepting, H. Yamase, and L. Zinni for illuminating discussions.
A part of the results
presented in this work was obtained by using the facilities of
the CCT-Rosario Computational Center, member of the High
Performance Computing National System (SNCAD, MincyT-
Argentina). A.~Greco acknowledges the Max-Planck-Institute for Solid State Research in Stuttgart for hospitality and financial support. X.W. is supported by the National Key R\&D Program of China (Grant No. 2023YFA1407300) and the National Natural Science Foundation of China (Grant No. 12447103).

\appendix

\section{Large-$N$ expansion based on the path integral for Hubbard operators. Extension to the bilayer $t$-$J_\parallel$-$J_\perp$}\label{app:fullformalism}

It is nontrivial to study the $t$-$J$ model because of the local constraint that prohibit double occupancy at any site.
In addition, the operators involved in the $t$-$J$ model are Hubbard operators $\hat{X}$ \cite{hubbard63}, which satisfy nonstandard commutation rules. In the language of Hubbard operators  
$\tilde{c}^\dag_{i \sigma}=\hat{X}_i^{\sigma 0}$, 
$\tilde{c}_{i \sigma}=\hat{X}_i^{0 \sigma}$, $S_i^+=\hat{X}_i^{\uparrow \downarrow}$, 
$S_i^-=\hat{X}_i^{\downarrow \uparrow}$, $n_i=\hat{X}_i^{\uparrow \uparrow}+\hat{X}_i^{\downarrow \downarrow}$, 
and $\hat{X}_i^{0 0}$ describes the number of doped holes; the $z$ component of the spin operator 
is described by $S_{i}^{z} = \frac{1}{2}(\hat{X}_{i}^{\uparrow \uparrow} -  \hat{X}_{i}^{\downarrow \downarrow})$.  
The operators $\hat{X}_i^{\sigma 0}$ and $\hat{X}_i^{0 \sigma}$ are called fermionlike, 
whereas the operators $\hat{X}_i^{\sigma \sigma'}$ and 
$\hat{X}_i^{00}$ are bosonlike. 

If $t_\perp=J_\perp=0$ the Hamiltonian Eq. (\ref{eq:H}) reduces to two 2D decoupled $t$-$J$ models which were treated in \cite{foussats04}.
See also the Appendix of Ref. \cite{yamase21} for a more detailed formulation of the path integral representation for Hubbard operators.

For clarifying ideas, here we focus on the case of finite $t_\perp$ and $J_\perp$, which on the other hand are the new contributions of interest for discussing pressurized nickelates. In terms of Hubbard operators the last two terms of Eq. (\ref{eq:H}) can be written as
\begin{eqnarray}
t_\perp \sum_{i,\sigma,\alpha}\; \hat{X}_{i,\alpha}^{\sigma 0}\hat{X}_{i,\bar{\alpha}}^{0\sigma} 
+ \frac{J_\perp}{4} \sum_{i, \sigma, \sigma', \alpha} (\hat{X}_{i,\alpha}^{\sigma \sigma'}\hat{X}_{i,\bar{\alpha}}^{\sigma' \sigma} - 
\hat{X}_{i,\alpha}^{\sigma \sigma} \hat{X}_{i,\bar{\alpha}}^{\sigma' \sigma'})\nonumber \\
\label{eq:last2} 
\end{eqnarray}
\noindent where $\alpha=1,2$ is an index that identifies each plane.

There are two major difficulties in the Hamiltonian Eq. (\ref{eq:H}): the complicated commutation rules of the Hubbard operators \cite{hubbard63} and the absence of a small perturbative parameter.
A popular method to handle the former difficulty is to use slave particles. 
For instance, in the slave-boson method \cite{kotliar88} 
the original fermionlike $\hat{X}^{0 \sigma}$ operator is written as 
a product of usual bosonic (holons) and fermionic (spinons) operators. 
This scheme introduces a gauge field, which requires a gauge fixing and the introduction of a Faddeev-Popov determinant \cite{leguillou95}.
Gauge fluctuations should be taken into account beyond mean-field theory and the slave particles 
need to be convoluted to form the original fermionic operator $\tilde{c}$.

We employ a large-$N$ technique based on the path integral representation for Hubbard operators \cite{foussats04,bejas12}.
In the large-$N$ approach we extend the spin projections from $\sigma = \uparrow,\downarrow$ to $p = 1, \cdots, N$, and rescale the Hamiltonian parameters to $t/N$, $t'/N$, $J_\parallel/N$, $t_\perp/N$, and $J_\perp/N$.
One of the advantages of this method is that it treats all possible charge excitations on an equal footing \cite{bejas12}.

Since our method works on the basis of path integral for Hubbard operators our fermions are just related to the Hubbard operators and not with the spinons of the slave-boson technique, and we do not have to deal with the Bose condensation of holons.
In addition, the fermionlike and bosonlike Hubbard operators are described by Grassmann and usual complex variables, respectively. Then, in the context of a path integral treatment, we remove the hat symbol of the Hubbard operators.

Following \cite{foussats04,yamase21}  we write 
\begin{equation}
{X}_{i,\alpha}^{\sigma \sigma'} = \frac{{X}_{i,\alpha}^{\sigma 0}
{X}_{i,\alpha}^{0 \sigma'}}{{X}_{i,\alpha}^{0 0}},  
\end{equation}

\noindent and write the fermionlike fields as follows:
\begin{eqnarray}
f^{\dagger}_{i p,\alpha} &=& \frac{1}{\sqrt{N \delta/2}} \; {X}_{i,\alpha}^{p 0} \,,\\
f_{i p,\alpha} &=& \frac{1}{\sqrt{N \delta/2}} \; {X}_{i,\alpha}^{0 p} \,.
\label{fermoper}
\end{eqnarray}
\noindent where $\delta$ is the hole doping rate away from half-filling. 
The fermion variable $f_{ip,\alpha}$ is proportional to ${X}_{i,\alpha}^{0 p}$ and should not be associated with the so-called spinon in the slave-boson method. 

The field ${X}_{i,\alpha}^{0 0}$ is written in terms of the fluctuations of the number of holes $\delta R_{i,\alpha}$ as
\begin{eqnarray}
&&{X}_{i,\alpha}^{0 0} = N \frac{\delta}{2}(1 + \delta R_{i,\alpha}) \,\label{hubbop} \, .
\end{eqnarray}

Making these changes the  exchange  interaction term ($J_\perp$-term) contains four fermion fields and the two bosons in the denominator.
They can be decoupled through a Hubbard-Stratonovich transformation by introducing the field $\Delta'_{i}$,
\be
\Delta'_{i} = \frac{J_\perp}{4N} \sum_{p} \frac{f^{\dag}_{i p, 1} f_{i p, 2}}{ \sqrt{(1 + \delta R_{i,1})
(1 + \delta R_{i,2})}} \,. 
\label{bond-field}
\ee
This field describes out-of-plane bond-charge fluctuations.

Therefore, Eq. (\ref{eq:last2}) is written as
\begin{align}
&t_\perp \frac{\delta}{2}\sum_{i,p,\alpha}
  f^{\dagger}_{i p,\alpha}f_{i p,\bar{\alpha}} 
+ \frac{2N}{J_\perp} \sum_{i} \Delta'^{\dagger}_{i}\;\Delta'_{i} \nonumber \\
-& \sum_{i,p} \left(
\frac{f^{\dagger}_{i p,1} f_{i p,2}}{ \sqrt{(1 +
 \delta R_{i,1}) (1 + \delta R_{i,2})}}\; \Delta'_{i} + {\rm H.c.}
 \right)
\end{align}

The out-of-plane bond field $\Delta'_{i}$ is parametrized as 
\be
\Delta'_i = \chi'(1+r_{\perp,i} + iA_{\perp,i})\,,
\label{eq:Deltap}
\ee
\noindent where $r_{\perp,i}$ and $A_{\perp,i}$ correspond to the real and imaginary parts of 
the out-of-plane bond-field fluctuations, respectively, and $\chi'$ is a static mean-field value. 
Finally, we expand the term $1/\sqrt{1+\delta R}$ in powers of $\delta R$, which generates various interactions between fermions and bosons.
The number of the interactions considered in the theory is controlled in powers of $1/N$. 

Putting together the above terms for the out-of-plane Hamiltonian Eq. (\ref{eq:last2}) and the in-plane terms \cite{foussats04,yamase21}, the effective theory for Eq. (\ref{eq:H}) is described in terms of fermions, bosons, and their interactions. 

In terms of the spinor $(f_{{\bf k}p,1},f_{{\bf k}p,2})$ we define a $2 \times 2$ electron Green's function which is $O(1)$
\begin{eqnarray}\label{eq:G0}
G^{(0)}_{\alpha \beta}(\vk,i\nu_n)= \left(
 \begin{array}{cc}
i\nu_n -\varepsilon^{\parallel}_{\vk} & -(t_\perp \delta/2-\chi')\\
   -(t_\perp \delta/2-\chi') & i\nu_n-\varepsilon^{\parallel}_{\vk}\
 \end{array}
\right)^{-1}, \nonumber \\
\end{eqnarray}
\noindent with 
\begin{align}{\label{eq:ek}}
\varepsilon_{\vk}^{\parallel} =& -2 \left( t \frac{\delta}{2}+\chi \right) (\cos k_{x}+\cos k_{y}) \nonumber \\
&+ 4t' \frac{\delta}{2} \cos k_{x} \cos k_{y} - \mu  \,,
\end{align}

For a given doping $\delta$, the chemical potential $\mu$, $\chi$, and $\chi'$ are determined self-consistently by solving
\begin{equation}{\label {chi}}
\chi= \frac{J_\parallel}{4N_s} \sum_{\vk,i\nu_n} (\cos k_x + \cos k_y) G^{(0)}_{11}(\vk,i\nu_n)\; , 
\end{equation}
\begin{equation}{\label {chi'}}
\chi'= \frac{J_\perp}{2N_s} \sum_{\vk,i\nu_n} G^{(0)}_{12}(\vk,i\nu_n)\; , 
\end{equation}
and 
\begin{equation}
 1 - \delta = \frac{2}{N_s} \sum_{\vk,i\nu_n} G^{(0)}_{11}(\vk,i\nu_n),
\end{equation}

In the above expressions $N_s$ and $\nu_n$ are the number of sites in each plane and a fermionic Matsubara frequency, respectively.
$G^{(0)}_{12}$ and $G^{(0)}_{11}$ are the elements $(1,2)$ and $(1,1)$ of the Green's function,  respectively.   

In the context of the $t$-$J_\parallel$-$J_\perp$ model using a path-integral representation for Hubbard operators a $14$-component bosonic field is defined as:
\begin{align}
\delta X^{a} = (&\delta R_1,\;\delta{\lambda_1},\;
                \delta R_2,\;\delta{\lambda_2},\;
                r_1^{x},\;r_1^{y},\; A_1^{x},\;A_1^{y},\; \nonumber \\
                & r_2^{x},\;r_2^{y},\; A_2^{x},\;A_2^{y},\;
                r_\perp,\; A_\perp)\, ,
\label{eq:boson-field}
\end{align}
where $\delta R_\alpha$ describes fluctuations of the number of holes at a given site (the site index is excluded for clarity) in the plane $\alpha$ and  it is related to on-site charge fluctuations in each plane, $\delta \lambda_\alpha$ is the fluctuation of the Lagrange multiplier introduced to enforce the constraint that prohibits the double occupancy at any site in each plane, and $r_\alpha^{x}$ and $r_\alpha^{y}$ ($A_\alpha^{x}$ and $A_\alpha^{y}$) describe fluctuations of the real (imaginary) part of the in-plane bond field coming from the $J_\parallel$-term.
$r_\perp$ and $A_\perp$ are the fluctuations of the real and imaginary part, respectively, of the out-of-plane bond field coming from the $J_\perp$-term.

At this point, and for completeness, it is important to mention how the local constraint
$X_{i,\alpha}^{0 0} + \sum_{p} X_{i,\alpha}^{p p} = N/2$
at each plane is treated.
These local constraints are then kept by introducing in the path integral the Lagrange multipliers $\lambda_{i,\alpha}$: 
\begin{align}
\delta&\left( X_{i,\alpha}^{0 0} + \sum_{p} X_{i,\alpha}^{p p} - N/2 \right)= \nonumber \\
&\int {\cal D} \lambda_{i,\alpha} 
\, {\rm exp}\left[
{\mathrm i} \lambda_{i,\alpha} \left( X_{i,\alpha}^{0 0} + \sum_{p} X_{i,\alpha}^{p p} - N/2 \right) 
\right] \,,
\end{align}
\noindent and writing $\lambda_{i,\alpha}$ in terms of
static mean-field values $\lambda_{0}$, which can be absorbed in the chemical potential, and dynamic fluctuations $\delta\lambda_{i,\alpha}$: 
\begin{align}
\lambda_{i,\alpha} = \lambda_{0} + \delta{\lambda_{i,\alpha}} \, .
\label{hubbop-lambda}
\end{align}

Following \cite{foussats04,bejas12} we obtain
\begin{align}
 [D^{(0)}_{ab}({\bf q},\mathrm{i}\omega_{n})]^{-1} &= 
 N \left(
 \begin{array}{ccccc}
   D^{(0)}_A & D^{(0)}_B & 0     & 0     & 0 \\
   D^{(0)}_B & D^{(0)}_A & 0     & 0     & 0 \\
        0 &      0 & D^{(0)}_C & 0     & 0 \\
        0 &      0 & 0     & D^{(0)}_C & 0 \\
        0 &      0 & 0     & 0     & D^{(0)}_D
 \end{array}
\right) \; ,
\label{eq:D0}
\end{align}
\noindent a $14 \times 14$ bare bosonic propagator $D^{(0)}_{ab}({\bf q},\mathrm{i}\omega_{n})$ associated with $\delta X^{a}$, where $\omega_n$ is a bosonic Matsubara frequency.
The matrices $D^{(0)}_{A\text{-}D}$ are
\begin{align}
 D^{(0)}_A &= \left(
 \begin{array}{cc}
  -\frac{\delta^2}{2} J(\vq) & \frac{\delta}{2} \\
  \frac{\delta}{2} & 0
 \end{array}
 \right)
\end{align}
\begin{align}
 D^{(0)}_B &= \left(
 \begin{array}{cc}
  -\frac{\delta^2}{2} J_{\perp}/2 & 0 \\
  0 & 0
 \end{array}
 \right)
\end{align}
\begin{align}
 D^{(0)}_C &= \left(
 \begin{array}{cccc}
  \frac{4\chi^2}{J_{\parallel}} & 0 & 0 & 0 \\
  0 & \frac{4\chi^2}{J_{\parallel}} & 0 & 0 \\
  0 & 0 & \frac{4\chi^2}{J_{\parallel}} & 0 \\
  0 & 0 & 0 & \frac{4\chi^2}{J_{\parallel}}
 \end{array}
 \right)
\end{align}
\begin{align}
 D^{(0)}_D &= \left(
 \begin{array}{cc}
  \frac{4\chi'^2}{J_{\perp}} & 0 \\
  0 & \frac{4\chi'^2}{J_{\perp}}
 \end{array}
 \right) \; ,
\end{align}
\noindent where $J({\bf q}) = (J_{\parallel}/2) (\cos q_x + \cos q_y)$. Note that $D^{(0)}_{ab}$ is $O(1/N)$.

The theory also gives three- and four-legs vertices.
The three-legs vertex $\Lambda_{\alpha\beta,a}$ represents the interaction of a fermion from the plane $\alpha$ that ends at the plane $\beta$ after interacting with a boson $\delta X^a$.
The nonzero components of this vertex are
\begin{align}\label{eq:threeleg1}
 &\Lambda_{\alpha\alpha,a}(\vk,\vq) =\nonumber \\
 &- \left[
 \frac{i\nu_n+i\nu'_n}{2} + \mu + 2\chi \sum_{\eta=x,y} \cos\left(k_\eta-\frac{q_\eta}{2}\right) \cos \frac{q_\eta}{2},\; 1,\; \right. \nonumber \\
 &-2\chi \cos\left(k_x-\frac{q_x}{2}\right),\; -2\chi \cos\left(k_y-\frac{q_y}{2}\right),\; \nonumber \\
 & \left. 2\chi \sin\left(k_x-\frac{q_x}{2}\right),\; 2\chi \sin\left(k_y-\frac{q_y}{2}\right)
 \right]
\end{align}
\noindent for each component $a = \delta R_\alpha$,$\delta \lambda_\alpha$,
$r^x_\alpha$, $r^y_\alpha$, $A^x_\alpha$, $A^y_\alpha$, and
\begin{align}\label{eq:threeleg2}
 \Lambda_{1\, 2,a}(\vk,\vq) =& - \left(  \frac{\chi'}{2}, \frac{\chi'}{2}, -\chi', -i\chi'  \right) \nonumber \\
 \Lambda_{2\, 1,a}(\vk,\vq) =& - \left(  \frac{\chi'}{2}, \frac{\chi'}{2}, -\chi',  i\chi'  \right)
\end{align}
\noindent for each component $a = \delta R_1$, $\delta R_2$, $r_{\perp}$, $A_{\perp}$.

The four-legs vertex $\Lambda_{\alpha\beta,ab}$ represents a fermion from the plane $\alpha$ that ends at the plane $\beta$ after interacting with the bosons $\delta X^a$ and $\delta X^b$.
The nonzero components of this vertex are
\begin{align}\label{eq:fourleg}
 \Lambda_{\alpha\alpha,ab} =& \left[
 \begin{array}{cc}
  F_{\bf q} & 1/2 \\
  1/2      & 0
 \end{array}
 \right]
\end{align}
\noindent for component $a,b = \delta R_\alpha$, $\delta \lambda_\alpha$, where
\begin{align}
 F_{\bf q} &= \frac{i\nu_n+i\nu'_n}{2} + \mu + \chi \sum_{\eta=x,y}
 \cos\left(k_\eta+\frac{q_\eta+q'_\eta}{2} \right)\nonumber \\
 &\times \left[
 \cos\left(\frac{q_\eta+q'_\eta}{2} \right) +
  \cos \frac{q_\eta}{2}  \cos \frac{q'_\eta}{2}
 \right] \; ,
 \nonumber 
\end{align}
\begin{align}
 \Lambda_{\alpha\alpha,ab} =& -\chi \left[
 \cos\left(k_x-\frac{q_x+q'_x}{2} \right), \, \cos\left(k_y-\frac{q_x+q'_y}{2} \right), \right. \, \nonumber \\
 &\left. -\sin\left(k_x-\frac{q_x+q'_x}{2} \right), \, -\sin\left(k_y-\frac{q_x+q'_y}{2} \right)
 \right]
\end{align}
\noindent for component $a = \delta R_\alpha$, and  $b = r^x_\alpha$, $r^y_\alpha$, $A^x_\alpha$, and $A^y_\alpha$, and
\begin{align}\label{eq:lam12ab}
 \Lambda_{1\,2,ab} =& -\frac{\chi'}{4}\left[
 \begin{array}{cccc}
  -3/2 & -1/2 & 1 & i \\
  -1/2 & -3/2 & 1 & i \\
   1   & 1    & 0 & 0 \\
   i   & i    & 0 & 0
 \end{array}
 \right]
\end{align}
\noindent for component $a,b = \delta R_1$, $\delta R_2$, $r_\perp$, $A_\perp$.
For the vertices $\Lambda_{2\,1,ab}$ the components involving $A_\perp$ change sign [right column and bottom row of Eq. (\ref{eq:lam12ab})]. Note that the vertices are $O(1)$.

\begin{figure}[t]
\centering
\includegraphics[width=8.1cm]{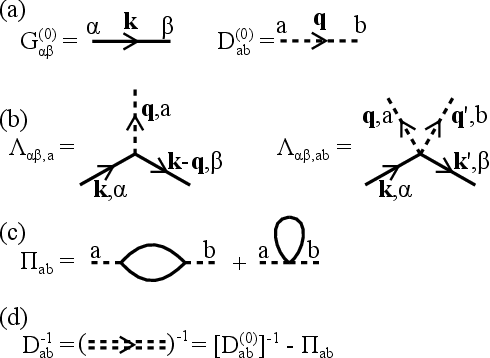}
\caption{(a) Bare fermionic $G^{(0)}_{\alpha \beta}$ and bosonic $D^{(0)}_{ab}$ propagators, solid and dashed lines, respectively. (b) $\Lambda_{\alpha\beta,a}$ and $\Lambda_{\alpha\beta,ab}$ are the three- and four-legs vertices, respectively.  (c) $\Pi_{ab}$ is the bosonic self-energy. (d) Double dashed line is the dressed bosonic propagator $D_{ab}$ beyond mean-field level.}
\label{fig:dyson}
\end{figure}

Using the propagators [Fig. \ref{fig:dyson}(a)] and vertices [Fig. \ref{fig:dyson}(b)] we can compute the bosonic self-energy $\Pi_{ab}({\bf q}, i\omega_n)$ [Fig. \ref{fig:dyson}(c)] which is used in the Dyson equation to compute the dressed bosonic propagator $D_{ab}({\bf q}, i\omega_n)$ [Fig. \ref{fig:dyson}(d)],
\begin{align}
 D^{-1}_{ab}({\bf q}, i\omega_n) &=\left[ D^{(0)}_{ab}({\bf q}, i\omega_n) \right]^{-1} - \Pi_{ab}({\bf q}, i\omega_n) \, .
 \label{eq:Dyson}
\end{align}

The order in $1/N$ of any calculated quantity is estimated by counting the number of propagators and vertices involved in the calculation \cite{foussats04}.

\section{The exchange $J_\perp$ subspace}\label{app:jperp}

As discussed in the main text, the relevant and new quantity is the out-of-plane exchange $J_\perp$.
Then here it is useful and instructive to focus on this sector, which is defined by the subspace generated by the elements $13$ and $14$ of the $\delta X^{a}$ [Eq. (\ref{eq:boson-field})], i.e., $r_\perp$ and $A_\perp$. 

In this subspace we have the following bare bosonic propagator and vertices:

a) The $2 \times 2$ bare bosonic propagator 
\begin{align}
 [D^{(0)}_{\textrm{z-BOP}}({\bf q},\mathrm{i}\omega_{n})]^{-1}= N \left(
 \begin{array}{cc}
  \frac{4\chi'^2}{J_{\perp}} & 0 \\
  0 & \frac{4\chi'^2}{J_{\perp}} 
 \end{array}
 \right) \;
\label{eq:Jperpsec}
\end{align}
\noindent associated to the out-of-plane bond-field fluctuations $r_\perp$ and $A_\perp$.

b) The three-legs vertices
\begin{align}\label{eq:Jperpvert}
 \Lambda_{1\, 2,a} =& \left( \chi',  i\chi' \right) \nonumber \\
 \Lambda_{2\, 1,a} =& \left( \chi', -i\chi' \right) 
 \end{align}
\noindent for $a = r_{\perp}, A_{\perp}$.

The $2 \times 2$ z-BOP bosonic self-energy is the sector 13-14 of $\Pi_{ab}$,
\begin{align}\label{eq:zbubble}
   \Pi_{ab}({\bf q}, i\omega_n) &= \sum_{k,\alpha,\beta}
                  \Lambda_{\alpha \bar{\alpha}, a}
                  G^{(0)}_{\alpha \beta}(k)
                  G^{(0)}_{\bar{\alpha} \bar{\beta}}(k-q)
                  \Lambda_{\beta \bar{\beta}, b} \, .
\end{align}
\noindent where $a, b = 13$, $14$.
Here we use the simplified notation $q$ meaning $\vq,i\omega_n$, $k$ meaning $\vk,i\nu_n$, and $k-q$ meaning $\vk-\vq,i\nu_n-i\omega_n$.
Using Eq. (\ref{eq:Jperpvert}) for the vertices we can explicitly write four contributions:
\begin{align}\label{eq:zbubbleexp}
   \Pi_{ab}(\vq,i\omega_n) = \sum_{k}
                  (\chi')^2 &\left[
                  \left(
                  \begin{array}{cc}
                   1 & i\\
                   i &-1
                  \end{array}
                  \right)
                  G^{(0)}_{1\,1}(k)
                  G^{(0)}_{2\,2}(k-q) + \right. \nonumber \\
                  &\left(
                  \begin{array}{cc}
                   1 &-i\\
                   i & 1
                  \end{array}
                  \right)
                  G^{(0)}_{1\,2}(k)
                  G^{(0)}_{2\,1}(k-q) + \nonumber \\
                  &\left(
                  \begin{array}{cc}
                   1 & i\\
                  -i & 1
                  \end{array}
                  \right)
                  G^{(0)}_{2\,1}(k)
                  G^{(0)}_{1\,2}(k-q) + \nonumber \\
                  &\left.
                  \left(
                  \begin{array}{cc}
                   1 &-i \\
                  -i &-1
                  \end{array}
                  \right)
                  G^{(0)}_{2\,2}(k)
                  G^{(0)}_{1\,1}(k-q)
                  \right] \, .
\end{align}
From the inverse of Eq. (\ref{eq:G0}) we can see that $G^{(0)}_{1\,1} = G^{(0)}_{2\,2}$, and $G^{(0)}_{1\,2} = G^{(0)}_{2\,1}$.
Using this we can write Eq. (\ref{eq:zbubbleexp}) as
\begin{align}\label{eq:zbubbleexp2}
   \Pi_{ab}(\vq,i\omega_n) = \sum_{k}
                  (\chi')^2 &\left[
                  \left(
                  \begin{array}{cc}
                   2 & 0\\
                   0 &-2
                  \end{array}
                  \right)
                  G^{(0)}_{1\,1}(k)
                  G^{(0)}_{1\,1}(k-q) + \right. \nonumber \\
                  &\left. \left(
                  \begin{array}{cc}
                   2 & 0\\
                   0 & 2
                  \end{array}
                  \right)
                  G^{(0)}_{1\,2}(k)
                  G^{(0)}_{1\,2}(k-q)
                  \right] \, ,
\end{align}
\noindent where it can be seen that the elements $\Pi_{13\,14} = \Pi_{14\,13} = 0$, and the explicit expressions for $\Pi_{13\,13}$ and  $\Pi_{14\,14}$ are
\begin{align}\label{eq:pies1314}
 \Pi_{13\,13} &= - (\chi')^2 \sum_{\bf k} \left( g^{--} + g^{++} \right) \\
 \Pi_{14\,14} &= - (\chi')^2 \sum_{\bf k} \left( g^{-+} + g^{+-} \right) \, ,
\end{align}
\noindent where
\begin{align}\label{eq:gab}
 g^{\alpha\beta} &=
 \frac{n_F(\varepsilon_{{\bf k}-{\bf q}}^{\alpha}) - n_F(\varepsilon_{{\bf k}}^{\beta})}
 {i\omega_n + \varepsilon_{{\bf k}-{\bf q}}^{\alpha} - \varepsilon_{{\bf k}}^{\beta}} \, , \nonumber \\
 \varepsilon_{{\bf k}}^{\pm} &= \varepsilon_{\bf k}^{||} \pm (t_\perp \delta/2-\chi') \, ,
\end{align}
\noindent and $n_F$ is the Fermi factor.

Then, in the $J_\perp$ subsector the Dyson equation reads as:
\begin{align}\label{eq:DzBOP}
&D_{\textrm{z-BOP}}^{-1}(\vq,i\omega_n)= \nonumber \\
&N \left(
 \begin{array}{cc}
\frac{4 {\chi'}^2}{J_\perp}-\Pi_{13\, 13}(\vq,i\omega_n)&0 \\
0&\frac{4 {\chi'}^2}{J_\perp}-\Pi_{14\, 14}(\vq,i\omega_n)
\end{array}
\right),
\end{align}
\noindent which defines the $2 \times 2$ dressed bosonic propagator beyond mean-field level [Eq. (\ref{eq:Jperpsec})].
Thus, the renormalized $D_{\textrm{z-BOP}}^{-1}$ results diagonal.

\section{Superconducting gap equations}\label{app:scgapeq}

For the calculation of the superconducting in-plane ($\Delta_{||}$) and out-of-plane ($\Delta_{\perp}$) gap we introduce the four-component Nambu spinor 
\begin{equation}
{\psi}^{\dagger}_{\bf k}= (f_{{\bf k} \uparrow,1}^{\dagger},\;f_{{\bf k} \uparrow,2}^{\dagger},\; f_{-{\bf k} \downarrow,1},\;f_{-{\bf k} \downarrow,2})\, ,
\label{spinorNambu}
\end{equation}
\noindent and the inverse of the $4 \times 4$ Nambu Green's function can be written as
\begin{align}\label{GN4x4}
&G^{-1}(\vk,i\nu_n)=\nonumber \\
&\left(
 \begin{array}{cccc}
i\nu_n -\varepsilon^{\parallel}_{\bf k} & -\varepsilon_{\perp} & -\Delta_\parallel(\vk) & -\Delta_\perp(\vk) \\
-\varepsilon_{\perp} & i\nu_n-\varepsilon^{\parallel}_{\bf k} & -\Delta_\perp(\vk) & -\Delta_\parallel(\vk) \\
-\Delta_\parallel(\vk) & -\Delta_\perp(\vk) & i\nu_n +\varepsilon^{\parallel}_{\bf k} & \varepsilon_{\perp} \\
-\Delta_\perp(\vk) & -\Delta_\parallel(\vk) & \varepsilon_{\perp} & i\nu_n +\varepsilon^{\parallel}_{\bf k} \
\end{array}
\right),
\end{align}
\noindent where $\varepsilon_{\perp} = (t_\perp \delta / 2 - \chi')$, $\Delta_\parallel(\vk) = \Delta_\parallel \gamma_d(\vk)$
and $\Delta_\perp(\vk) = \Delta_\perp$, $i\nu_n$ is a fermionic Matsubara frequency, and $\varepsilon^{\parallel}_{\bf k}$ is the band dispersion in the plane [Eq. (\ref{eq:ek})].

Then, the gap equations for $\Delta_\parallel$ and $\Delta_\perp$ are
\begin{equation}\label{eq:gappar}
\Delta_\parallel=-J_\parallel \sum_{\vk, i\nu_n} \gamma_d(\vk) G_{13}(\vk,i\nu_n) 
\end{equation}
\noindent and 
\begin{equation}\label{eq:gapper}
\Delta_\perp=-\frac{J_\perp}{4} \sum_{\vk,i\nu_n} G_{14}(\vk,i\nu_n), 
\end{equation}
\noindent and must be solved self-consistently.
In Eq. (\ref{eq:gappar}) $\gamma_d({\bf k}) = \frac{1}{2}(\cos k_x - \cos k_y)$.

\section{z-BOP susceptibilities and gap equation for the z-BOP gap}\label{app:zBOPsus}

The z-BOP susceptibilities are defined projecting $D_{\textrm{z-BOP}}$ on the eigenvectors $(1,0)$ and $(0,1)$, and are given respectively by
\begin{align}
\label{eq:sus13}
\chi^{r_\perp}_{\textrm{z-BOP}}(\vq,i\omega_n)=& \left[
\frac{4 {\chi'}^2}{J_\perp}-\Pi_{13 \,13}(\vq,i\omega_n)
\right]^{-1} \, , \\
\label{eq:sus14}
\chi^{A_\perp}_{\textrm{z-BOP}}(\vq,i\omega_n)=& \left[
\frac{4 {\chi'}^2}{J_\perp}-\Pi_{14 \,14}(\vq,i\omega_n)
\right]^{-1} \, .
\end{align}

For the calculation of the z-BOP gap $\phi$ we introduce the four-component Nambu spinor 
\begin{equation}
{\psi}^{\dagger}_{\bf k}= (f_{{\bf k} \sigma,1}^{\dagger},\;f_{{\bf k} \sigma,2}^{\dagger},\; f_{{\bf k}+{\bf q}_c \sigma,1}^{\dagger},\;f_{{\bf k}+{\bf q}_c \sigma,2}^{\dagger})\, ,
\label{spinorzBOP}
\end{equation}
\noindent and the inverse of the $4 \times 4$ Nambu Green's function for each spin can be written as
\begin{align}\label{zBOP4x4}
&G^{-1}(\vk,i\nu_n)= \nonumber \\
&\left(
 \begin{array}{cccc}
i\nu_n -\varepsilon^{\parallel}_{\bf k} & -\varepsilon_{\perp} & 0& -\phi \\
-\varepsilon_{\perp} & i\nu_n-\varepsilon^{\parallel}_{\bf k} & -\phi & 0 \\
0 & -\phi & i\nu_n -\varepsilon^{\parallel}_{{\bf k}+ {\bf q}_c} & -\varepsilon_{\perp} \\
-\phi & 0 & -\varepsilon_{\perp} & i\nu_n -\varepsilon^{\parallel}_{{\bf k}+{\bf q}_c} \
\end{array}
\right),
\end{align}
\noindent where ${\bf q}_c$ is the ordered momentum of the z-BOP.

Then, the gap equation for $\phi$ is 
\begin{equation}\label{eq:phi4x4}
\phi=-\frac{J_{\perp}}{2} \sum_{\vk, i\nu_n} G_{14}(\vk,i\nu_n) \, .
\end{equation}
This order parameter is associated with the interplane bond order $\langle f_{{\bf k} \sigma,1}^{\dagger} f_{{\bf k}+{\bf q}_c \sigma,2} \rangle$. Thus, as mentioned in Sec. \ref{sec:zBOP}, a finite value of $\phi$ indicates a modulation of the interlayer hopping with wave vector ${\bf q}_c$.

Note that Eq. (\ref{zBOP4x4}) is valid for a commensurate momentum such as ${\bf q}_c = (\pi,\pi)$.
For an incommensurate momentum one should use a bigger matrix with the additional symmetry related momenta.
As the values we used for ${\bf q}_c$ are close to $(\pi,\pi)$ and the $T_c^{\rm z-BOP}$ versus doping obtained when the z-BOP gap $\phi$ becomes nonzero is in good agreement with the one obtained from the susceptibilities [see Fig. \ref{just}(a)], we think that using Eq. (\ref{zBOP4x4}) with an incommensurate momentum ${\bf q}_c$ is a good first approximation.

\section{Competition between z-BOP and superconductivity}\label{app:competition}
We introduce the eight-spinor field
\begin{align}
{\psi}^{\dagger}_{\bf k}= (& f_{{\bf k} \uparrow,1}^{\dagger},\;f_{{\bf k} \uparrow,2}^{\dagger},\; f_{-{\bf k} \downarrow,1},\;f_{-{\bf k} \downarrow,2}, \nonumber \\
&f_{{\bf k}+{\bf q}_c \uparrow,1}^{\dagger},\;f_{{\bf k}+{\bf q}_c \uparrow,2}^{\dagger},\; f_{-{\bf k}-{\bf q_c} \downarrow,1},\;f_{-{\bf k}-{\bf q}_c \downarrow,2})\,
\label{spinorcomp}
\end{align}
\noindent where ${\bf q}_c$ is the ordered momentum of the z-BOP.
The inverse of the $8 \times 8$ Green's function can be written as
\begin{eqnarray}\label{GN}
G^{-1}(\vk,i\nu_n)=
\left(
\begin{array}{cc}
A&B\\
B&C\
\end{array}
\right),
\end{eqnarray}
\noindent with $A$, $B$, and $C$ the following $4 \times 4$ matrices 
\begin{align}\label{GNA}
&A= \nonumber \\
&\left(
\begin{array}{cccc}
i\nu_n -\varepsilon^{\parallel}_{\bf k} & -\varepsilon_{\perp} & 0 &-\Delta_\perp\\
-\varepsilon_{\perp} & i\nu_n-\varepsilon^{\parallel}_{\bf k} & -\Delta_\perp& 0 \\
0 & -\Delta_\perp& i\nu_n +\varepsilon^{\parallel}_{\bf k} & \varepsilon_{\perp}\\
-\Delta_\perp & 0 & \varepsilon_{\perp} & i\nu_n + \varepsilon^{\parallel}_{\bf k}
\end{array}
\right),
\end{align}

\begin{eqnarray}\label{GNB}
B=
\left(
\begin{array}{cccc}
0&-\phi&0&0\\
-\phi&0&0&0\\
0&0&0&\phi\\
0&0&\phi&0\
\end{array}
\right),
\end{eqnarray}
\noindent and $C$ has the form of $A$ with $\vk+{\bf q}_c$ instead of $\vk$. $\Delta_\perp$ and $\phi$ are the out-of-plane superconducting and z-BOP gap, respectively. The gap equations for $\Delta_\perp$ and $\phi$ are
\begin{equation}
\Delta_\perp=-\frac{J_\perp}{4} \sum_{\vk,i\nu_n} G_{14}(\vk,i\nu_n), 
\label{eq:Deltaperp}
\end{equation}
\noindent and 
\begin{equation}
\phi=-\frac{J_\perp}{2} \sum_{\vk,i\nu_n} G_{16}(\vk,i\nu_n) 
\label{eq:phi}
\end{equation}
\noindent and must be solved self-consistently.

\bibliography{main_sub_AN}

\begin{thebibliography}{97}
\bibitem [{mis()}]{misc-factor2a}%
  \BibitemOpen
  \href@noop {} {}\bibinfo {note} {{Note that we have to use 1eV for $t$
  because the $1/N$ scaling in the large-$N$ expansion
  \cite{foussats04}.}}\BibitemShut {Stop}%
\end{thebibliography}

\end{document}